\documentclass[11pt]{article}
\usepackage{amsmath}
\newcommand{\eq}[1]{(\ref{#1})}
\newcommand{\be}{\begin{equation}}
\newcommand{\ee}{\end{equation}}
\newcommand{\bea}{\begin{eqnarray}}
\newcommand{\eea}{\end{eqnarray}}

\newcommand{\vs}[1]{\vspace{#1 mm}}
\newcommand{\hs}[1]{\hspace{#1 mm}}

\def\a{\alpha}

\let\bm=\bibitem
\def\nn{\nonumber}

\begin{document}
	
%

\centerline{\bf Superhairs on the  Branes of  D =11 Supergravity}
	\vs{10}
\centerline{Ali Kaya }
\vs{5}
\centerline{Department of Physics and Astronomy, Texas A\&M University,}
\centerline{College Station, Texas 77843, USA.}
\vs{15}
\begin{abstract}
It is shown that membrane and fivebrane of $\mathrm{D}=11$ supergravity theory can support nongauge, linearized spin-$3 / 2$ superhairs. Supercharges associated with these fields are calculated. We also generalize the solutions to some overlapping cases and discuss possible implications of their existence.
\end{abstract}

\section{Introduction} 

Supergravity theories are invariant under local supersymmetry transformations which, for asymptotically flat solutions, gives rise to new fermionic Noether currents. Spinor charges associated with these currents can be calculated as surface integrals at spatial infinity. In the canonical Hamiltonian formulation, after gauge fixing, these charges together with the total linear and angular momenta, which can also be expressed as surface integrals, obey the global supersymmetry algebra \cite{1,2}. A purely bosonic solution of a supergravity theory has zero supercharge. When the theory is linearized with respect to fermionic fields, a supersymmetry transformation leaves the bosonic background invariant but, in general generates fermions which are solutions of the linearized field equations. This configuration can be viewed as an approximate solution. Moreover, when the parameter of the transformation approaches a constant value at infinity, long-range fields are generated that have appropriate falloff properties to carry supercharges \cite{3} .

However, supersymmetry is a certain local gauge symmetry and fermion fields which are obtained by transformations tend to identity at infinity must be thought to be "pure gauge". If this is the case, then the approximate solution obtained by supersymmetry must be identified with the bosonic background (Gauss's law). The long-range fields mentioned above are not pure gauge in that sense (since in that case the transformation is not the identity at infinity). Recently, starting with the work \cite{n1}  large gauge transformations  have been shown to imply the existence of a rich asymptotic symmetry structure of gauge theories that connects soft theorems \cite{n3}  and memory effect \cite{n4}  (for a pedagogical review see \cite{n2}).  Not surprisingly, these results can be extended to supergravity theories giving rise to asymptotic supersymmetry algebras \cite{n5} 

In any case,  one might search for fermionic perturbations of bosonic backgrounds that cannot be obtained by any supersymmetry transformation (from now on the terms "nongauge" and "nontrivial" will only be used for such fields to distinguish them from the fields obtained by supersymmetry). This was studied for black holes of $\mathrm{O}(2)$ extended supergravity theory in four dimensions. In the papers \cite{4,5,6,7}  it was shown that among the class of Kerr-Newman black holes only the extreme Reissner-Nordström family admits nongauge fermionic hairs carrying nonzero supercharges. It is remarkable that these are the only known black hole solutions in four dimensions that can accept supercharge as a new parameter. There are other studies examining the superhair problem for  black hole solutions in various supergravity theories, for a recent study see \cite{y1}. 

$\mathrm{D}=11$ supergravity theory admits various extended black hole solutions i.e. black p-branes \cite{8}. These solutions have the structure of an extended object surrounded by an event horizon, are asymptotically flat and characterized by two parameters; the mass per unit p-volume and the charge associated with the four-form field. Only the extreme black p-branes, which are obtained when the mass and the charge saturates a Bogomolny type of bound, preserve some fraction of supersymmetry. Among the extreme solutions, p=2 membrane and p=5 fivebrane are particularly interesting. All other extreme members can be interpreted as various intersections of them \cite{9,10} (for a review see also \cite{11}). Furthermore, they are playing a crucial role in the derivation of various string dualities from the eleven-dimensional M-theory (for introduction to M-theory see \cite{12,13,14} ).

As black p-branes are natural higher-dimensional generalizations of the four-dimensional black holes, one may wonder how the superhair problem works out for them. This is actually more than a pure academic interest because of the status of string/M theory as a framework of quantum gravity. For example, the original proposal of the AdS/CFT  correspondence  uses the supergravity solution of $N$ coincident D3 branes of type IIB string theory \cite{mal}. Similarly, the near horizon decoupling  limits of the intersecting brane solutions yield various AdS/CFT dualities with a lower number of supersymmetries and brane solutions in D=11 are used to deduce AdS/CFT dualities involving M-theory  \cite{mal}. These obviously show the importance of the (supersymmetric) p-brane solutions and this motivates one to search for their new unexplored properties. 

In this paper we show that, like the four-dimensional extreme Reissner- Nordström black holes, the extreme membrane and fivebrane backgrounds in 11 dimensions can also support nongauge superhairs. During the investigation we do not try to solve the linearized spin $3 / 2$ field equations (up to supersymmetry transformations) to find the complete set of nontrivial solutions, but we use a simple limiting procedure, described in \cite{15}, to obtain a single mode on each background. As we will see these modes have the correct falloff properties as they are approaching to spatial infinity yielding nonzero supercharges. Inspired by the structure of nontrivial fields on the membrane and fivebrane backgrounds, we also write a general ansatz for the overlapping solutions. By using this ansatz, we find similar spin $3 / 2$ fields on the two membranes overlapping at a point and on the two fivebranes overlapping at a three brane. We will discuss some possible implications of these superhairs in the conclusion.

\section{Limiting Procedure}

The field variables of $\mathrm{D}=11$ supergravity theory are the metric $g_{a b}$, the four form field $F_{a b c d}$ and the spin $3 / 2$ gravitino field $\psi_{a}$. When the theory is linearized with respect to gravitinos, field equations reduce to a Rarita-Schwinger equation on a fixed bosonic background, namely \footnote{The $\Gamma$ matrices are purely imaginary and the signature of the metric is $(+,-,-\ldots . .-)$}
\bea
&&	R_{a b}=-\frac{\kappa^{2}}{36} g_{a b}\left(F^{c d} F_{c d}\right)+\frac{\kappa^{2}}{3} F_{a c d e} F_{b}^{. c d e}, \label{1}\\
&&	\nabla_{m} F^{m a b c}+\frac{\kappa}{4 \times 144} \epsilon^{m n p q e f g h a b c} F_{m n p q} F_{e f g h}=0, \label{2}\\
&&	\kappa \Gamma^{a b c} D_{b} \psi_{c}=0, \label{3}
\eea
where $\kappa$ is the gravitational coupling constant, $R_{a b}$ is the Ricci tensor $\nabla_{a}$ is the usual covariant derivative of the metric $g_{a b}$ and $D_{a}$ is the supercovariant derivative given by:
\be\label{4}
D_{a}=\frac{1}{\kappa} \nabla_{a}+\frac{i}{144}\left(\Gamma_{a}^{. b c d e}-8 \Gamma^{c d e} \delta_{a}^{b}\right) F_{b c d e} .
\ee
Note that the presence of gravitinos does not affect the background geometry, specifically its contribution to the torsion vanishes and supercovariant derivative $D_{a}$ depends only on the bosonic fields. In this approximation to the full theory, supersymmetry transformations leave $g_{a b}$ and $F_{a b c d}$ invariant but $\psi_{a}$ shifts as
\be\label{5}
\psi_{a} \rightarrow \psi_{a}+D_{a} \epsilon,
\ee
where $\epsilon$ is the parameter of the transformation. It can be shown, with the help of equations \eq{1} and \eq{2}, that if $\psi_{a}$ is a solution of (3) then $\psi_{a}+D_{a} \epsilon$ is also a solution. This ensures the supersymmetry invariance of the linearized theory. We will call $\psi_{a}$ is trivial or gauge generated if it is equal to $D_{a} \epsilon$ for some $\epsilon$ since then it can set to be zero by a supersymmetry transformation.

There is a simple method of obtaining nontrivial spin $3 / 2$ fields from covariantly constant spinors \cite{15}. Let us describe this for membrane and fivebrane we are interested in. The key point is that they are members of a two parameter family of solutions characterized by two positive reals $r_{+}$ and $r_{-}$ which are related to the mass and charge. Generic solutions $\left(r_{+}>r_{-}\right)$ break all the supersymmetries which means that they have no Killing spinors. Let $\epsilon\left(r_{+}, r_{-}\right)$ be a spinor field on one of these backgrounds which is arbitrary unless in the limit $r_{+} \rightarrow r_{-}$it becomes the Killing spinor of the extreme solution (in what fallows, we keep $r_{-}$ fixed and vary $r_{+}$). Then, the following limit
\be\label{6}
\psi_{a} \equiv \lim _{r_{+} \rightarrow r_{-}} \frac{1}{r_{+}-r_{-}} D_{a}\left(r_{+}, r_{-}\right) \epsilon\left(r_{+}, r_{-}\right)
\ee
exists and $\psi_{a}$ satisfies the Rarita-Schwinger equation \eq{3} on the extreme background, where $D_{a}\left(r_{+}, r_{-}\right)$ is the supercovariant derivative of the generic background. Moreover, gravitino mode obtained in this way is in general nontrivial. It is easy to see that $\psi_{a}$ is a solution if it exists (take the limit after inserting it into equation \eq{3}). Thus, let us show the existence of the limit. 

Near extremality, the supercovariant derivative $D_{a}\left(r_{+}, r_{-}\right)$ and the spinor $\epsilon\left(r_{+}, r_{-}\right)$ can be expanded around $r_{+}=r_{-}$ as
\bea
&&	D_{a}\left(r_{+}, r_{-}\right)=D_{a}^{0}+\lambda W_{a}+O\left(\lambda^{2}\right), \label{7}\\
&&	\epsilon\left(r_{+}, r_{-}\right)=\epsilon_{0}+\lambda \beta+O\left(\lambda^{2}\right)\label{8}
\eea
where $\lambda=r_{+}-r_{-}$,  $D_{a}^{0} \equiv D_{a}\left(r_{+}=r_{-}\right)$, $\epsilon_{0} \equiv \epsilon\left(r_{+}=r_{-}\right)$, $W_{a}$ is an operator and $\beta$ is a certain spinor. Now $\epsilon_{0}$ is, by construction, the Killing spinor of the extreme solution i.e. $D_{a}^{0} \epsilon_{0}=0$. Then \eq{6} becomes
\be
\psi_{a}=W_{a} \epsilon_{0}+D_{a}^{0} \beta
\ee
which proves the existence of the limit. The term $D_{a}^{0} \beta$ is manifestly pure gauge. Different choices of $\epsilon\left(r_{+}, r_{-}\right)$ correspond to different choices of $\beta$ which in turn gives gauge-equivalent fields. A necessary condition for the gravitino to be gauge generated on the extreme background requires
\be\label{10}
D_{[a}^{0} \psi_{b]}=\frac{1}{2} \Omega_{a b} \tilde{\epsilon},
\ee
where $\tilde{\epsilon}$ is a certain gauge spinor and $\Omega_{a b}$ denotes the supercurvature matrix defined by
\be\label{11}
\left[D_{a}^{0}, D_{b}^{0}\right]=\Omega_{a b} .
\ee
curvature matrix depends on the bosonic field variables and at this moment we do not need an explicit expression for it. In general, the condition \eq{10} is not satisfied by the fields obtained in the limiting procedure.

\section{Superhairs on the membrane and fivebrane} 

Let us first consider the generic black-membrane solution which is given by
\bea
&&d s^{2}=\Delta_{+} \Delta_{-}^{-1 / 3} d t^{2}-\Delta_{+}^{-1} \Delta_{-}^{-1} d r^{2}-r^{2} d \Omega_{7}^{2}-\Delta_{-}^{2 / 3}\left(d x_{1}^{2}+d x_{2}^{2}\right), \label{12}\\
&&	F=\frac{q_{e}}{r^{7}} d t \wedge d r \wedge d x_{1} \wedge d x_{2}\label{13}
\eea
where $\Delta_{\pm}=1-\left(r_{\pm}^{6} / r^{6}\right)$, $d \Omega_{7}^{2}$ is the line element on the 7-sphere $S^{7}$ and $q_{e}=18 r_{+}^{3} r_{-}^{3}$. This is an electrically-charged solution with total charge $q_{e} \Omega_{7}$ where $\Omega_{7}$ is the area of $S^{7}$. The surface $r=r_{+}$ is a regular event horizon and $r=r_{-}$ is a singular surface hidden behind this horizon. The charge parameter $q_{e}$ and the mass per unit 2-volume $\mu$ satisfies $\mu \geq q_{e}$. The extreme membrane solution, in the warped product form, is obtained by setting $r_{0} \equiv r_{+}=r_{-}$
\bea
&& d s^{2}=\Delta^{2 / 3}\left(d t^{2}-d x_{1}^{2}-d x_{2}^{2}\right)-\Delta^{-2} d r^{2}-r^{2} d \Omega_{7}^{2} ,\label{14}\\
&&	F=\frac{q_{e}}{r^{7}} d t \wedge d r \wedge d x_{1} \wedge d x_{2} \label{15} 
\eea
where now, $\Delta=1-r_{0}^{6} / r^{6}$ and $q_{e}=18 r_{0}^{6}$. In this case, the surfaces at $r=r_{+}$ and $r=r_{-}$ coalesce at $r=r_{0}$ and become a regular event horizon. Also, the above bound between $\mu$ and $q_{e}$ is saturated; $\mu=q_{e}$. The limit \eq{6} gives (after a suitable supersymmetry transformation and up to normalization) the one-form gravitino field
\be\label{16} 
\psi=\psi_{\hat{0}} E^{\hat{0}}+\psi_{\hat{i}} E^{\hat{i}}
\ee
where
\bea
&&	\psi_{\hat{0}}=\frac{2}{r^{7}} \Delta^{-1} \Gamma^{\hat{0} \hat{r}} \epsilon_{0} ,\label{17}\\
&&	\psi_{\hat{i}}=\frac{1}{r^{7}} \Delta^{-1} \Gamma^{\hat{i} \hat{r}} \epsilon_{0}, \label{18}
\eea
$\epsilon_{0}$ is the Killing spinor of the extreme membrane, and hatted quantities and spinors refer to the basis one-forms $E^{\hat{0}}=\Delta^{1 / 3} d t$,  $E^{\hat{i}}=\Delta^{1 / 3} d x^{i}$, $E^{\hat{r}}=\Delta^{-1} d r$, $E^{\hat{\theta}}=r e^{\theta}$ where $e^{\theta}$ denotes the basis one-forms of the 7-sphere. The Killing spinor $\epsilon_{0}$ obeys $\Gamma^{\hat{r} \hat{\theta}_{1} \ldots \hat{\theta}_{7}} \epsilon_{0}=-\epsilon_{0}$.

On the other hand, fields of a black fivebrane may be written as
\bea
&&	d s^{2}=\Delta_{+} \Delta_{-}^{-2 / 3} d t^{2}-\Delta_{+}^{-1} \Delta_{-}^{-1} d r^{2}-r^{2} d \Omega_{4}^{2}-\Delta_{-}^{1 / 3}\left(d x_{1}^{2}+\ldots +d x_{5}^{2}\right), \hs{7}\label{19}\\
&&	F=q_{m} \epsilon_{4},\label{20}
\eea
where $\Delta_{\pm}=1-r_{\pm}^{3} / r^{3}$, $d \Omega_{4}^{2}$ is the line element and $\epsilon_{4}$ is the volume form of $S^{4}$ and $q_{m}=9 r_{+}^{3 / 2} r_{-}^{3 / 2}$. This is a magnetically-charged solution with total charge $q_{m} \Omega_{4}$ where $\Omega_{4}$ is the area of $S^{4}$. Again, $r=r_{+}$ is a regular event horizon and $r=r_{-}$ is a singular surface. Similar to the black-membrane case there is a bound for the mass per unit 5-volume $\mu$ given by $\mu \geq q_{m}$. The extreme fivebrane is obtained by setting $r_{0} \equiv r_{+}=r_{-}$
\bea
	d s^{2}=\Delta^{1 / 3}\left(d t^{2}-d x_{1}^{2}-\ldots-d x_{5}^{2}\right)-\Delta^{-2} d r^{2}-r^{2} d \Omega_{4}^{2} \label{21}\\
	F=q_{m} \epsilon_{4}\label{22}
\eea
where now $\Delta=1-r_{0}^{3} / r^{3}$, $q_{m}=9 r_{0}^{3}$. Like for the extreme membrane, $r=r_{0}$ becomes a regular event horizon and the bound between the mass and charge is saturated $\mu=q_{m}$. This time the limit (6) gives (again after a suitable supersymmetry transformation and up to normalization) the gravitino one-form field
\be
\psi=\psi_{\hat{0}} E^{\hat{0}}+\psi_{\hat{i}} E^{\hat{i}} \label{23}
\ee
where
\bea
\psi_{0}=\frac{5}{r^{4}} \Delta^{-1} \Gamma^{\hat{0} \hat{r}} \epsilon_{0} ,\label{24}\\
\psi_{\hat{i}}=\frac{1}{r^{4}} \Delta^{-1} \Gamma^{\hat{i} \hat{r}} \epsilon_{0},\label{25}
\eea
$\epsilon_{0}$ is the Killing spinor of extreme fivebrane, the basis one-forms are  $E^{\hat{0}}=\Delta^{1 / 6} d t$, $E^{\hat{i}}=\Delta^{1 / 6} d x^{i}$, $E^{\hat{r}}=\Delta^{-1} d r$ and $E^{\hat{\theta}}=r e^{\theta}$ where $e^{\theta}$ denotes the basis one-forms of the 4-sphere. The Killing spinor $\epsilon_{0}$ obeys $\Gamma^{\hat{r} \hat{\theta}_{1} \ldots \hat{\theta}_{4}} \epsilon_{0}=-i \epsilon_{0}$.

The one-form spin-$3 / 2$ fields given in \eq{16} and \eq{23} are static solutions of \eq{3} on the extreme membrane and fivebrane backgrounds respectively. To show that they are nontrivial fields one may use the integrability condition \eq{10}. For this it is enough to consider $(0, i)$ components of the supercurvature tensor. A simple calculation shows that $\Omega_{0 i}=0$, both for the membrane and fivebrane. However from the spin-$3 / 2$ fields given in \eq{17}-\eq{18} and \eq{24}-\eq{25} one can find for the membrane
\be\label{26}
D_{[0}^{0} \psi_{i]}=-\Delta^{-1 / 3} \frac{r_{0}^{6}}{r^{14}} \Gamma^{\hat{0} \hat{i}} \epsilon_{0}
\ee
and for the fivebrane
\be\label{27}
D_{[0}^{0} \psi_{i]}=-\Delta^{-2 / 3} \frac{r_{0}^{3}}{r^{8}} \Gamma^{\hat{\hat{i}} \hat{\epsilon}} \epsilon_{0}
\ee
Therefore condition \eq{10}  is not satisfied for our fields proving that they are nongauge.

Let us make a few comments about the behavior of the gravitino modes at spatial infinity and at the event horizon. Supercharges (per p-volumes) associated with a membrane and a fivebrane are calculated as surface integrals on 7- and 4-spheres at infinity surrounding the extended objects. As the transverse radial coordinate $r \rightarrow \infty$ , one has $\psi \sim 1 / r^{7}$ on the membrane and $\psi \sim 1 / r^{4}$ on the fivebrane. As we will show these are the required falloff properties for the fields to carry supercharges. On the other hand, the spin-$3 / 2$ hairs \eq{17}-\eq{18} and \eq{24}-\eq{25} seem to be singular on the event horizon, when $r=r_{0}$. However these are simply coordinate singularities; the coordinates that are used to express the metrics given in \eq{14} and \eq{21} are ill defined at $r=r_{0}$ (yet the geometries are perfectly regular as in the Schwarzschild black hole \cite{16,17}). Moreover, the tangent space bases chosen to express the spinor fields  are not regular at $r=r_{0}$ either (note the bases given after \eq{18} and \eq{25}). There is an indirect argument showing that the gravitino modes obtained by the limiting procedure are indeed regular at $r=r_{0}$. Clearly, the term $D_{a}\left(r_{+}, r_{-}\right) \epsilon\left(r_{+}, r_{-}\right)$ in \eq{6} is regular at $r=r_{+}$ since the generic black-brane solution is regular at the event horizon. Let us denote an arbitrary entry of $D_{a}\left(r_{+}, r_{-}\right) \epsilon\left(r_{+}, r_{-}\right)$, with respect to a regular tangent space basis, by $f\left(r_{+}, r_{-}\right)$. When $r_{+}=r_{-}$, we know that $f\left(r_{+}, r_{-}\right)=0$ (remember the definition of $\epsilon\left(r_{+}, r_{-}\right)$). We have previously shown that near extremality $f\left(r_{+}, r_{-}\right)$ can be expanded  as $f\left(r_{+}, r_{-}\right)=\left(r_{+}-r_{-}\right) g$ where now $g=g(r)$ is the corresponding entry of $\psi_{a}$ with respect to the same regular basis on the extreme solution. As $r \rightarrow r_{0}$, $g(r)$ cannot diverge since otherwise we can let $r_{+} \rightarrow r_{-}$ with a suitable rate to obtain a nonzero $f$ which in turn contradicts the fact that $f=0$ when $r_{+}=r_{-}$. This shows that the gravitino field must be regular at the horizon.

For a branelike configuration, the spin 3/2 supercharge per unit p-volume can be constructed from the behavior of the gravitino at spatial infinity \cite{18}
\be\label{28}
Q=\int_{\partial E} \Gamma^{a b c} \psi_{c}\, d \Sigma_{a b}
\ee
where $E$ is the transverse space to the extended object. For our interest $\partial E$ is a 7-sphere and a 4-sphere at infinity surrounding membrane and fivebrane, respectively at an instant of time. Thus $d \Sigma_{a b}=d \Sigma_{\hat{0} \hat{r}}$. Remembering also the fact that the area element of an n-sphere $S^{n}$ goes like $r^{n}$, the supercharges associated with \eq{16} and \eq{23} can be calculated as (up to multiplicative constants which depends on the normalization of the gravitino modes)
\be\label{29}
Q \sim \Gamma^{\hat{0}} \eta
\ee
where $\eta$ is a constant chiral spinor satisfying $\Gamma^{(8)} \eta=-\eta$ for the membrane and $\Gamma^{(5)} \eta=-i \eta$ for the fivebrane ($\eta$ is the asymptotic value of the Killing spinor thus obey the same chirality conditions) where $\Gamma^{(8)}$ and $\Gamma^{(5)}$ are the completely antisymmetric combinations of the 8- and 5-transverse gamma matrices. Thus, for each case, $Q$ spans a 16-dimensional subspace of spinor space. We have mentioned that nonzero supercharges can also be obtained by supersymmetry transformations with constant parameters at infinity. For the extreme membrane and fivebrane, since they preserve $1 / 2$ fraction of supersymmetry, supercharges associated with fields obtained by large supersymmetry transformations also span a 16-dimensional subspace. This time the parameter of the transformation must obey the opposite chirality conditions obeyed by Killing spinors. Supercharges corresponding to these fields can be calculated as
\be\label{30}
Q_{\text {susy }} \sim \Gamma^{\hat{0}} \eta_{\text {susy }},
\ee
where now the constant spinor $\eta_{\text {susy }}$ (the value of the supersymmetry parameter at infinity) obeys $\Gamma^{(8)} \eta_{\text {susy }}=+\eta_{\text {susy }}$ and $\Gamma^{(5)} \eta_{\text {susy }}=+i \eta_{\text {susy. }}$. Thus $Q$ and $Q_{\text {susy }}$ apparently belong to the different subspaces and they together span the 32-dimensional spinor space.

Extreme membrane and fivebrane solutions allow generalizations to multibrane configurations. It turns out that, the gravitino modes obtained on the single membrane and fivebrane can also be generalized to become solutions on these backgrounds. In constructing these generalizations it is convenient to use Euclidean coordinates for the transverse spaces. For the extreme membrane one can define a new radial coordinate $R$ with $R^{6}=r^{6}-r_{0}^{6}$ and introduce an eight-dimensional Euclidean space with coordinates $y^{\alpha}$ and $R^{2}=y^{\alpha} y^{\alpha}$. In this coordinate system \eq{14}, \eq{15} and spin $3 / 2$ field \eq{16} looks like
\bea
&&	d s^{2}=U^{-2 / 3}\left(d t^{2}-d x_{1}^{2}-d x_{2}^{2}\right)-U^{1 / 3}\left(d y^{\alpha} d y^{\alpha}\right), \label{31}\\
&&	F=3 U^{-2} d t \wedge d U \wedge d x_{1} \wedge d x_{2}, \label{32}\\
&&	\psi=2\left(\frac{1}{R^{7}} U^{-1 / 2} \Gamma^{\hat{0} \hat{r}} \epsilon_{0}\right) d t+\left(\frac{1}{R^{7}} U^{-1 / 2} \Gamma^{\hat{i} \hat{r}} \epsilon_{0}\right) d x^{i},\label{33}
\eea
where $U=1+\left(r_{0}^{6} / R^{6}\right)$, basis one-forms are chosen as $E^{\hat{0}}=U^{-1 / 3} d t$, $E^{\hat{i}}=U^{-1 / 3} d x^{i}$, $E^{\hat{\alpha}}=U^{1 / 6} d y^{\alpha}$ and $\Gamma^{\hat{r}}=y^{\alpha} \Gamma^{\hat{\alpha}} / R$ with $\Gamma^{\hat{r}} \Gamma^{\hat{r}}=-I$. For the extreme fivebrane, let us define $R^{3}=r^{3}-r_{0}^{3}$ and introduce 5 Euclidean coordinates $y^{\alpha}$ on the transverse space with $R^{2}=y^{\alpha} y^{\alpha}$. This time, the fields given in \eq{21}, \eq{22} and \eq{23} can be reexpressed in this new coordinate chart as
\bea
&&d s^{2}=U^{-1 / 3}\left(d t^{2}-d x_{1}^{2} \cdots-d x_{5}^{2}\right)-U^{2 / 3} d y \cdot d y, \label{34}\\
&&	F=3(* d U),\label{35} \\
&&	\psi=5\left(\frac{1}{R^{4}} U^{-1 / 2} \Gamma^{\hat{0} \hat{r}} \epsilon_{0}\right) d t+\left(\frac{1}{R^{4}} U^{-1 / 2} \Gamma^{\hat{i} \hat{r}} \epsilon_{0}\right) d x^{i},\label{36}
\eea
where $U=1+r_{0}^{3} / R^{3}$, $*$ is the Hodge dual on the transverse Euclidean space and spinors are defined with respect to the basis $E^{\hat{0}}=U^{-1 / 6} d t$, $E^{\hat{i}}=U^{-1 / 6} d x^{i}$ and $E^{\hat{\alpha}}=U^{1 / 3} d y^{\alpha}$. Again $\Gamma^{\hat{r}}=y^{\alpha} \Gamma^{\hat{\alpha}} / R$ with $\Gamma^{\hat{r}} \Gamma^{\hat{r}}=-I$. 

For $U$ being more general harmonic functions ($U=1+\sum c_{i} /\left(R-R_{i}\right)^{7}$ for the membrane and $U=1+\sum c_{i} /\left(R-R_{i}\right)^{4}$ for the fivebrane representing parallel membranes and fivebranes located at $R_{i}$, respectively) it can be checked that \eq{33} and \eq{36} remain to be solutions of \eq{3} on the multi-membrane and multi-fivebrane configurations. This result suggests that more general overlapping solutions may also support nontrivial superhairs although the limiting procedure, in the form described in this paper, cannot be applied to these cases.

\section{Generalization to the overlapping cases}

To show the existence of nongauge spin-$3 / 2$ fields on overlapping branes we use an ansatz  inspired by the structure of the spin-$3 / 2$ fields given in \eq{33} and \eq{36}.  For this analysis, we note that  the field equation for the linearized spin-$3 / 2$ field on a bosonic background  can be rewritten as
 \be\label{37}
\Gamma^{b}\left(D_{a} \psi_{b}-D_{b} \psi_{a}\right)=0.
\ee
This can be seen by using the gamma matrix identity
\be\label{38}
\Gamma^{a} \Gamma^{b c}=\Gamma^{a b c}+g^{a b} \Gamma^{c}-g^{a c} \Gamma^{b} .
\ee
Assuming \eq{37}, the identity \eq{38} implies
\be\label{39}
\Gamma^{a} \Gamma^{b c}\left(D_{b} \psi_{c}-D_{c} \psi_{b}\right)=\Gamma^{a b c}\left(D_{b} \psi_{c}-D_{c} \psi_{b}\right).
\ee
Contracting this with $\Gamma_{a}$ gives
\be\label{40}
\Gamma^{b c}\left(D_{b} \psi_{c}-D_{c} \psi_{b}\right)=0,
\ee
which also implies by \eq{39} that $\psi_{a}$ satisfies \eq{3}. One can repeat the same steps to show that \eq{3} implies \eq{37} and thus they are equivalent. In verifying the gravitino field equations we use \eq{37}, which saves us from considerable gamma-matrix algebra. 

Let us now consider overlapping membranes at a point, which have the coordinates $\left(t, x^{1}, x^{2}\right)$ and $\left(t, x^{3}, x^{4}\right)$. The fields of that configuration are given by \cite{10} 
\bea
&&d s^{2}=\left(U_{1} U_{2}\right)^{-2 / 3} d t^{2} -U_{1}^{-2 / 3} U_{2}^{1 / 3}\left(d x^{i}\right)^{2}\nn \\
	&&\hs{7}-U_{1}^{1 / 3} U_{2}^{-2 / 3}\left(d x^{m}\right)^{2}-\left(U_{1} U_{2}\right)^{1 / 3}\left(d y^{\alpha}\right)^{2}, \\
	&&F_{t 12 \alpha}= \frac{1}{2} \frac{\partial_{\alpha} U_{1}}{U_{1}^{2}}, \hs{5}F_{t 34 \alpha}=\frac{1}{2} \frac{\partial_{\alpha} U_{2}}{U_{2}^{2}} .
\eea
where $(i=1,2)$, $(m=3,4)$ , $(\alpha=5 \ldots 10)$, and $U_{1}$ and $U_{2}$ are harmonic functions of the coordinates $y^{\alpha}$. This solution preserves $1 / 4$ th of the supersymmetries of the vacuum. For the spin $3 / 2$ field we consider the following ansatz:
\bea
&&	\psi_{t} =A(y) \Gamma^{\hat{t} \hat{r}} \epsilon_{0} ,\label{44}\\
&&	\psi_{i} =B_{1}(y) \Gamma^{\hat{i} \hat{r}} \epsilon_{0}, \\
&&	\psi_{m} =B_{2}(y) \Gamma^{\hat{m} \hat{r}} \epsilon_{0} ,\\
&&	\psi_{\alpha} =C(y) \Gamma^{\hat{\alpha} \hat{r}} \epsilon_{0}\label{47} 
\eea
where $\epsilon_{0}$ is the Killing spinor and hatted quantities refers to the obvious tangent space basis. Note that, unlike the configurations on the parallel membranes and fivebranes, \eq{33} and \eq{36}, we allow transverse components for the gravitino. After a tedious algebra, the field equation \eq{37} can  be seen to fix the unknowns in terms of the harmonic functions as follows:
\bea
&&	A =\frac{2}{R^{5}}\left(U_{1} U_{2}\right)^{-1 / 2}, \\
&&	B_{1} =\frac{1}{R^{5}} U_{1}^{-1 / 2} ,\\
&&	B_{2} =\frac{1}{R^{5}} U_{2}^{-1 / 2}, \\
&&	C =-\frac{1}{2 R^{5}}.
\eea

Similarly, one can also consider overlapping fivebranes over a three-brane solution. Assuming  that the fivebranes are lying on the  $\left(t, x^{1}, x^{2}, x^{3}, x^{4}, x^{5}\right)$ and $\left(t, x^{1}, x^{2}, x^{3}, x^{6}, x^{7}\right)$ planes, the metric and the four-form field of such a configuration are given by \cite{10}
\bea
&&d s^{2}=\left(U_{1} U_{2}\right)^{-1 / 3}\left(d t^{2}-\left(d x^{a}\right)^{2}\right)-U_{1}^{-1 / 3} U_{2}^{2 / 3}\left(d x^{i}\right)^{2}\nn \\
&&\hs{7}-U_{1}^{2 / 3} U_{2}^{-1 / 3}\left(d x^{m}\right)^{2}-\left(U_{1} U_{2}\right)^{2 / 3}\left(d y^{\alpha}\right)^{2} ,\\
&&F_{12 \alpha \beta}=\frac{1}{2} \epsilon_{\alpha \beta \gamma} \partial_{\gamma} U_{2}, \hs{5} F_{34 \alpha \beta}=\frac{1}{2} \epsilon_{\alpha \beta \gamma} \partial_{\gamma} U_{1} 
\eea
where $(a=1,2,3)$, $(i=4,5)$, $(m=6,7)$, $(\alpha, \beta=8,9,10)$, and $U_{1}$ and $U_{2}$ are harmonic functions of $y^{\alpha}$. This solution  preserves $1 / 4$ th of the supersymmetries of the theory. By choosing an ansazt similar to \eq{44}-\eq{47}, a linearized spin $3 / 2$ mode on the fivebranes overlapping on a three brane can be determined as
\bea
&&\psi_{t} =\frac{5}{R^{2}}\left(U_{1} U_{2}\right)^{-1 / 2} \Gamma^{\hat{\hat{r}} \hat{r}} \epsilon_{0}, \label{55} \\
&& \psi_{a} =\frac{1}{R^{2}}\left(U_{1} U_{2}\right)^{-1 / 2} \Gamma^{\hat{a} \hat{r}} \epsilon_{0} ,\\
&&\psi_{i} =\frac{1}{R^{2}}\left(U_{1}\right)^{-1 / 2} \Gamma^{\hat{i} \hat{r}} \epsilon_{0} ,\\
&&\psi_{m} =\frac{1}{R^{2}}\left(U_{2}\right)^{-1 / 2} \Gamma^{\hat{m} \hat{r}} \epsilon_{0} ,\\
&&\psi_{\alpha} =\frac{-2}{R^{2}} \Gamma^{\hat{\alpha} \hat{r}} \epsilon_{0}\label{59}
\eea
where $\epsilon_{0}$ is the Killing spinor of the background. To see that spin $3 / 2$ fields \eq{44}-\eq{47} and \eq{55}-\eq{59} are nongauge we try to find an $\tilde{\epsilon}$ satisfying $D_{a} \tilde{\epsilon}=\psi_{a}$.  We are able to show that such an $\tilde{\epsilon}$ does not exist for each case by checking the integrability condition \eq{26} or \eq{27}. It is  natural to expect that nontrivial spin $3 / 2$ fields having a similar structure can be found on all overlapping solutions. Note that the components of the gravitino along a certain brane direction is multiplied by the inverse square root of the corresponding harmonic function. Unlike paralel membrane and fivebrane cases,  the spin $3 / 2$ field has non-vanishing components along the transverse directions on the overlapping solutions. The fields also have suitable falloff properties to support supercharges.

\section{Conclusions} 

What are the possible physical implications of the existence of nongauge superhairs on the brane solutions of D=11 supergravity? At the classical level,  the backgrounds with superhairs correspond to approximate solutions of the full supergravity theory involving fermionic fields. With additional conditions like the normalizability of the (linearized) modes, the whole set of solutions are expected to form supermultiplets of the underlying supersymmetry algebra, see e.g. \cite{y2}. It is not obvious how to directly measure a classical fermionic field since by definition it is an anti-commuting variable having c-number and soul parts. Yet one may imagine an interpretation similar to fermionic condensates or one can search their indirect impact on measurable physical quantities like the paths of test particles. 

In a semi-classical description, the eigenmodes of linearized field operators around a classical solution can be interpreted as the excitations of the corresponding solitonic object. In particular, the zero modes govern the low energy soliton dynamics. Generally, there are zero modes associated with the  moduli of the background solution like the translational zero modes or the fermionic modes obtained by large gauge transformations. These modes can be implemented as  world-volume fields and they  form supersymmetric multiplets. 

The gravitino zero modes obtained in this paper look like isolated excitations that are not related to the moduli.  Moreover,  they cannot be implemented as  world-volume fields since the constant parameters "characterizing" the modes cannot be imposed to be functions of the brane coordinates as these are the parameters of the unbroken global supersymmetries which must be kept constant from the world-volume theory point of view. Therefore, the existence of a nontrivial gravitino zero mode on a brane-like soliton must simply imply the existence of a zero energy fermionic ground state of the world-volume theory. Note that on the world volume there is already a balance between the fermionic and the bosonic degrees of freedom (usual zero modes) without the nontrivial zero modes. However supersymmetry allows their existence since they only contribute to the zero energy ground states. In general one may have arbitrary number of zero energy fermionic and bosonic states; the difference usually being equal to a topological invariant related to the index of an operator. These states form trivial one dimensional supersymmetry multiplets i.e they are singlets. It would be interesting to study the implications of these modes in the context of AdS/CFT duality and asymptotic symmetries. 

Finally, it has been conjectured that in quantum gravity there must be a general upper bound on the strength of gravity relative to gauge interactions, which is called the weak gravity conjecture (WGC) \cite{wg1}. The conjecture implies, among other things, that there are constraints on the mass-charge spectrum in four dimensional theories with gravity and U(1) gauge fields. Specifically, one would expect to find light particles whose mass to charge ratios are smaller than the corresponding ratio for the  macroscopic extremal black holes, hence allowing the extremal black holes to decay. Remarkably,  nontrivial arguments based on black hole thermodynamics support the WGC; as it turns out, the higher derivative corrections to Einstein-Maxwell and Einstein-Maxwell-dilaton theories modify the extremality condition for black holes so that the large extremal black holes are unstable  decaying to smaller extremal black holes as predicted by WGC \cite{wg2,wg3}. These results show that any classical computation based on background supergravity solutions should be carefully examined in the context of WGC and decaying instabilities. Not surprisingly,  black p-brane solutions are also modified by higher derivative terms and  in the context of string theory these can be expressed as $\a'$ corrections, see e.g. \cite{corr1,corr2}. Nevertheless, all brane solutions studied in this paper preserve some fraction of the supersymmetries of D=11 supergravity theory and  supersymmetric solutions enjoy various stability properties  thanks to the remarkable features of supersymmetry. For example, while the generic black p-brane solutions suffer from the well known linearized perturbation instability \cite{gl1}, the supersymmetric ones do not \cite{gl2}. The stability of the  supersymmetric states in the context of WGC has been indicated in \cite{corr3}. 

{\bf Acknowledgements:}  Dedicated to the memory of Rahmi Güven; a great mathematical physicist, a true mentor and an exemplary person.

\end{document}